\documentclass[12pt]{article}
\usepackage[margin=1.25in]{geometry}
\usepackage[utf8]{inputenc}
\usepackage{amsmath, amsthm, amssymb, amsfonts}
\usepackage{import}
\usepackage[numbers, comma]{natbib}
\usepackage{fullpage}
\usepackage{authblk}
\usepackage{apxproof}
\usepackage{import}
\usepackage{mathtools}
\usepackage{tikz}
\usepackage{float}
\usepackage{enumitem}
\usepackage{subfig}

\usetikzlibrary{calc,patterns,angles,shapes, positioning, intersections, quotes}

\newtheoremrep{thm}{Theorem}
\newtheoremrep{lem}{Lemma}
\newtheoremrep{prop}{Proposition}

\usepackage{booktabs}      
\usepackage{multirow} 
\renewcommand{\arraystretch}{1.5}  
\usepackage{makecell}      

\usepackage{graphicx}

\theoremstyle{definition}

\theoremstyle{example}

\newtheorem{assumption}{Assumption}

\usepackage[colorlinks=false,pagebackref=true, hidelinks]{hyperref}

\newcommand{\R}{\mathbb{R}}

\def\th{\theta}
\def\Th{\Theta}
\def\l{\lambda}
\def\B{\mathrm{B}}
\def\E{\mathbb{E}}
\def\thl{\underline{\theta}}
\def\thh{\overline{\theta}}

\allowdisplaybreaks

\date{\today\\\small{\href{https://goelsumit.com/files/contests_grading.pdf}{(Link to latest version)}}}

\begin{document}
\title{Optimal grading contests\thanks{I am grateful to Federico Echenique, Omer Tamuz, Thomas Palfrey, Jingfeng Lu, Wade Hann-Caruthers, and seminar audiences at the 24th ACM Conference on Economics and Computation (EC'23), 3rd Durham Economic Theory Conference, 8th Annual Conference on Contests: Theory and Evidence (2022), the Winter school at Delhi School of Economics (2022), and the 17th Annual conference at Indian Statistical Institute Delhi (2022) for helpful comments and suggestions. An earlier version of this paper circulated under the
title ``Prizes and effort in contests with private information". }
}

\author{Sumit Goel\thanks{New York University Abu Dhabi; sumitgoel58@gmail.com; 0000-0003-3266-9035}
}


\maketitle  

\begin{abstract}

We study the design of effort-maximizing grading schemes between agents with private abilities. Assuming agents derive value from the information their grade reveals about their ability, we find that more informative grading schemes induce more competitive contests. In the contest framework, we investigate the effect of manipulating individual prizes and increasing competition on expected effort, identifying conditions on ability distributions and cost functions under which these transformations may encourage or discourage effort. Our results suggest that more informative grading schemes encourage effort when agents of moderate ability are highly likely, and discourage effort when such agents are unlikely.

\end{abstract} 
\section{Introduction}

Contests are situations in which agents compete by investing costly resources to win valuable prizes. In many such situations, the prizes are not monetary and instead, take the form of grades which may be valuable to the agents because of the information they reveal about their private abilities. Some examples include classroom settings or massive open online courses, where students compete for better grades which they can use to signal their productivity to the market and get potentially higher wages. In such environments, the designer can choose how much information to reveal about the performance of the participating agents with their choice of a grading scheme, which might also influence the effort that the agents exert towards getting better grades. In this paper, we study how the informativeness of grading schemes influences the effort exerted by agents and identify the grading schemes that maximize expected effort.\\

We study this problem in a model where agents have private information about their abilities, measured by their marginal cost of effort. The designer, who can rank agents in terms of the effort they exert, commits to a grading scheme which describes the extent to which different ranks will be pooled together under a common grade. The grade obtained by an agent serves as a signal about their private ability, and the resulting posterior amounts to a wage offer in the market. Interpreting this wage offer as a prize, one corresponding to each rank, every grading scheme thus induces a unique contest in the classical sense between the agents. Moreover, under this modeling of grading schemes as signaling instruments, we show that more informative grading schemes induce more competitive contests, and in general, the aggregate sum of prizes in contests induced by grading schemes is constant. Thus, the question of how informativeness of grading schemes influences effort reduces to one of how competition influences effort in contests, and the design problem reduces to one of feasibly distributing a fixed budget among the different prizes.\\

This motivates our investigation of how manipulating prizes and increasing competition influences expected effort in contests where agents have private abilities. In this environment with linear costs, \citet{moldovanu2001optimal} show that increasing competition by transferring value to the first prize always encourages effort, and consequently, a budget-constrained designer would optimally allocate the entire budget to the first prize (if feasible) to maximize expected effort. While this optimality of the most competitive winner-takes-all contest suggests that increasing competition might generally encourage effort, we show that this is not necessarily the case. More precisely, we identify sufficient conditions on ability distributions under which increasing any intermediate prize, or increasing competition by transferring value between intermediate prizes, may encourage or discourage effort. These conditions suggest that the effect of increasing any intermediate prize is determined by the relative likelihood of productive and unproductive agents, while the effect of increasing competition is determined by the relative likelihood of moderately productive and extreme (high or low) productivity agents.\footnote{The case where the marginal costs of effort are uniformly distributed on $[0,1]$ turns out to be a useful reference point. With linear costs, the expected effort is simply the difference of the first and the last prize, and thus, the transformations involving intermediate prizes actually have no effect on expected effort.} \\

For general costs, we show that the effect these transformations have on equilibrium effort can sometimes be informed by their effect on equilibrium effort under linear costs. Intuitively, with general costs, one can equivalently model the contest as one in which agents directly choose effort costs instead of effort, and thus, the effect these transformations have on effort under linear costs can more generally be interpreted as the effect these transformations have on effort costs under arbitrary cost functions. This effect on effort cost has implications for the effect on effort itself when the cost function is concave or convex. In particular, we show that increasing intermediate prizes encourages effort under convex costs if it increases effort costs (encourages effort under linear costs), and it discourages effort under concave costs if it decreases effort costs (discourages effort under linear costs). For the effect of increasing competition, we are able to similarly infer the effect on effort from those on effort costs, under the additional condition that the transformation encourage effort from the most productive agent.\\

We use these results to solve for effort-maximizing grading schemes. As noted above, the problem is equivalent to distributing a fixed a budget across different prizes, but the restriction to distributions feasible under grading schemes imply bounds on the values that can be allocated to various prizes, and in particular, renders the winner-takes-all contest infeasible. Under linear costs, while the effort-maximizing grading rule still allocates the maximum possible value to the first prize (by uniquely identifying the best-performing agent), the optimal distribution of the left-over budget among the remaining prizes (the full structure of the grading scheme)  depends on how increasing competition effects effort, and thus, on the likelihood of moderately productive agents. If moderately productive agents are likely, competition encourages effort and the optimal grading scheme is highly informative, revealing exactly the rank obtained by some of the best-performing agents while pooling the remaining bottom ranked agents together. And if moderately productive agents are unlikely, competition discourages effort and the optimal grading scheme is relatively coarse as it uniquely identifies the best-performing agent, but pools all the remaining agents with at most two more different grades. 

\subsection*{Literature review}

There is a vast literature studying the classical contest design problem of distributing a fixed budget across various prizes. In models where agents have private abilities, which is the focus of this paper, \citet*{moldovanu2001optimal} show that the winner-takes-all contest is optimal under linear and concave costs, while \citet*{zhang2024optimal} identifies sufficient conditions under which it is optimal with convex costs as well. In comparison, in the complete information environment, \citet*{barut1998symmetric} establish an invariance result under linear costs, while \citet*{fang2020turning} show more generally that increasing competition encourages effort under concave costs, while discouraging effort under convex costs.\footnote{\citet*{glazer1988optimal} solve the budget distribution problem in some special cases of private ability and complete information environments, highlighting the distinct nature of the optimal contests in them.} \citet*{baranski2024contest} present a unifying approach by studying finite type-space domains, establishing the strict optimality of winner-takes-all contest under linear and concave costs even with the slightest uncertainty about abilities. \citet*{olszewski2016large, olszewski2020performance} study large contests, and show that it is generally optimal to award multiple prizes of descending sizes under convex costs.\footnote{The literature has also explored some variants of the design problem, investigating the role of general architectures (\citet*{moldovanu2006contest}), allocation rules (\citet*{letina2023optimal}), negative prizes (\citet*{liu2018optimal}), costly entry (\citet*{liu2019optimal}), and homogeneous prizes (\citet*{liu2017optimal}). Surveys of literature in contest theory can be found in \citet*{sisak2009multiple,corchon2007theory,vojnovic2015contest,konrad2009strategy, chowdhury2023heterogeneity, fu2019contests, bevia2024contests}.} Our paper investigates the more general problem of how increasing competition influences effort when agents have private information about their abilities, showing in particular that the effects are a bit more nuanced in comparison to the complete information case.\\

There is also some related literature studying how grading schemes influence effort. In closely related work, \citet*{moldovanu2007contests} and \citet*{immorlica2015social} also show that effort-maximizing grading schemes may involve coarse or fine partitioning depending upon the underlying distribution of abilities. In particular, \citet*{moldovanu2007contests} study a model similar to ours with relative grading schemes (grades can only depend on rank), but assume linear costs and use a very different approach leading to different conditions on ability distributions. \citet*{immorlica2015social} also assume linear costs, and allow for absolute grading schemes (grades can depend on absolute effort), showing that the optimal grading scheme generally reveals the exact ranks of agents above a certain effort threshold. In recent work, \citet*{krishna2024pareto} take the agents' perspective, and illustrate how pooling ranks together can be Pareto improving for students in contests for college admissions.  In a complete information setting, \citet*{dubey2010grading} find that absolute grading schemes dominate relative grading schemes for enticing effort from students.\footnote{Some other related work explores the design of absolute grading schemes in single-agent environments  (\citet*{rayo2013monopolistic, zubrickas2015optimal, rodina2016inducing, onuchic2023conveying}), and the structure of equilibrium grading schemes in models of university competition (\citet*{popov2013university, boleslavsky2015grading, ostrovsky2010information, chan2007signaling}), offering rationales for the general trend towards information suppression, grade inflation, and deteriorating grading standards in universities. \citet*{brownback2018classroom} and \citet*{butcher2023making} study empirically how grading schemes influence effort in the classroom.} Our paper contributes to this literature by studying the design of relative grading schemes as information disclosure policies, and our approach generates insights about the more fundamental question of how more informative grading schemes influence effort in general environments.\\


The paper proceeds as follows. In Section 2, we present the model of a contest and note some useful facts. Section 3 characterizes the symmetric Bayes-Nash equilibrium and analyzes the effect of increasing prizes and competition on expected effort. In Section 4, we introduce and discuss our application to the design of grading schemes as information disclosure policies. Section 5 concludes. All proofs are relegated to the appendix.  

\section{Model}

There is a set $N=\{1, 2 ,\dots, n\}$ of risk-neutral agents. Each agent $j\in N$ has a private type $\th_j$ (which captures its marginal cost of effort), drawn independently from type-space $\Th=[\thl, \thh]$ according to distribution $F: \Th \to [0,1]$ so that $\Pr[\theta_j\leq \theta]=F(\theta)$. In contrast to previous literature, we allow for the possibility that $\thl=0$, but assume that in such a case, extremely productive agents are not too likely. 

\begin{assumption}
\label{ass:geniusrare}
The distribution of abilities $F:[\thl, \thh] \to [0,1]$ is such that  
\begin{enumerate}
    \item it admits a differentiable density function  $f(\th)=F'(\th)$ for $\th\in (\thl, \thh)$,  
    \item $\lim_{t \to 0} th(t)=0$, where $h(t)=\frac{t}{F^{-1}(t)}$ for $t\in (0,1)$.
\end{enumerate}
\end{assumption} 

Notice that if $\thl>0$, Assumption \ref{ass:geniusrare} only requires that $F(.)$ is twice-differentiable, as the second condition is trivially satisfied. When $\thl=0$, the second condition can be equivalently written as $\lim_{\th \to 0} \frac{F^2(\th)}{\th} =0$, capturing the requirement that extremely productive agents are somewhat unlikely. For instance, with $\Theta=[0,1]$ and $F(\theta)=\theta^p$, $F(.)$ satisfies Assumption \ref{ass:geniusrare} if and only if $p>\frac{1}{2}$. \\

There is a designer who designs a contest $v=(v_1, \dots, v_n)$ with $v_1\geq v_2\geq \dots \geq v_n$. Given the contest $v$, all agents simultaneously choose their effort. The agents are ranked according to their effort and awarded the corresponding prizes, with ties broken uniformly at random. If agent $j\in N$ exerts effort $x_j\geq 0$ and wins prize $v_i$, its payoff is 
$$v_i-\theta_jc(x_j),$$
where $c:\R_+\to \R_+$ is a strictly increasing and differentiable cost function with $c(0)=0$ and $\lim_{x\to \infty} c(x)=\infty$. We will let $g=c^{-1}$.\\

Given $F(.)$ and $c(.)$, a contest $v$ defines a Bayesian game between the $n$ agents. We will focus on the symmetric Bayes-Nash equilibrium of this contest game. This is a strategy profile where all $n$ agents use the same strategy $X_v:\Th \to \R_+$ so that for any agent $j\in N$ with type $\th_j$, choosing effort $X_v(\th_j)$ leads to a payoff at least as high as the payoff from choosing any other effort level, given that all remaining $n-1$ agents are also using the strategy $X_v(.)$.\\

From the designer's perspective, we assume that it cares about maximizing expected effort. Formally, the designer prefers a contest $v$ over $v'$ if and only if $\mathbb{E}[X_{v}(\theta)] \geq \mathbb{E}[X_{v'}(\theta)]$, where $X_{v}(.)$ represents the symmetric Bayes-Nash equilibrium induced by contest $v$.

\subsection*{Notation and Facts}
In our analysis, we make use of some notation and facts that we note here. \\

We denote by $p_i(t)$ the probability that a random variable $Y \sim Bin (n-1,t)$ takes the value $i-1$. That is, for $i\in \{1, \dots, n\}$,  $$p_i(t)={n-1 \choose i-1} t^{i-1}(1-t)^{n-i}.$$ 
Note that $p_i(t)$ can be interpreted as the probability that an agent loses to exactly $i-1$ agents out of $n-1$ agents, and thus the probability that the agent wins prize $i$, when it loses to any arbitrary agent with probability $t$. The following is a useful fact about $p_i(t)$.
\begin{lemrep}
\label{lem:calculation}
For any $i \in \{1, \dots, n\}$ and $k\in \R$ such that $i+k>1$, 
$$\int_0^1 t^kp_i'(t)dt =p_i(1)-k{n-1 \choose i-1}\B(i+k-1,n-i+1)$$ where $\B(a,b)=\int_0^1 t^{a-1}(1-t)^{b-1}dt$ represents the Beta function.
\end{lemrep}
\begin{proof}
For any $i \in \{1, \dots, n\}$ and $k>1-i$,
\begin{align*}
   \int_0^1 t^kp_i'(t)dt &= t^kp_i(t)|^1_0-\int_0^1 kt^{k-1}p_i(t)dt\\
   &=p_i(1)- k{n-1 \choose i-1}\int_0^1  t^{i+k-2}(1-t)^{n-i}dt\\
   &=p_i(1)-k{n-1 \choose i-1}\B(i+k-1,n-i+1)
\end{align*}

\end{proof}

Some of our findings depend on the distribution of abilities $F(.)$, and in particular, depend on whether $h(.)$ is monotone increasing or decreasing or if it is concave or convex. We note here some conditions on $F(.)$ which are potentially easier to verify and sufficient for $h(.)$ to have these properties.

\begin{lemrep}
\label{lem:conditions}
Suppose $F:[\thl, \thh] \to [0,1]$ is a distribution that satisfies Assumption \ref{ass:geniusrare}.

\begin{enumerate}
\item If $f(\th)$ is increasing on $(\thl, \thh)$, then $h(t)$ is increasing on $(0, 1)$.
\item If $f(\th)$ is decreasing on $(\thl, \thh)$ and $\thl=0$, then $h(t)$ is decreasing on $(0, 1)$.
\item $\frac{f(\th)\th^2}{F^2(\th)}$ is decreasing on $(\thl, \thh)$ $\iff$ $h(t)$ is concave on $(0, 1)$.
\item $\frac{f(\th)\th^2}{F^2(\th)}$ is increasing on $(\thl, \thh)$ and $\thl=0$ $\iff$ $h(t)$ is convex on $(0, 1)$.
\end{enumerate}
\end{lemrep}
\begin{proof}
Let $\th(t)=F^{-1}(t)$ so that $h(t)=\dfrac{F(\th(t))}{\th(t)}$. It follows that
\begin{align*}
   h'(t) &=\dfrac{\th f(\th) -F(\th)}{\th^2} \theta'(t)\\
   &=\dfrac{\th f(\th) -F(\th)}{\th^2 f(\th)}.
\end{align*}
Differentiating again, we get
\begin{align*}
h''(t)&=\dfrac{\th ^2f(\th)(\th f'(\th )+f(\th )-f(\th ))-(\th f(\th )-F(\th ))(2\th f(\th )+\th ^2f'(\th ))}{\th^4 f^2(\th )}\th '(t)\\
&=\dfrac{\th ^2f(\th )f'(\th )-(\th f(\th )-F(\th ))(2 f(\th )+\th f'(\th ))}{\th ^3 f^3(\th )}\\
&=\dfrac{1}{\th^3 f(\th )^3}\left[F(\th)(2f(\th)+\th f'(\th))-2\th f^2(\th)\right].
\end{align*}

Now we can prove the claims in order.
\begin{enumerate}
\item If $f(\th)$ is increasing on $(\thl, \thh)$, then $$F(\th)=\int_{\thl}^\th f(t)dt \leq \int_{\thl}^\th f(\th)dt=(\th-\thl)f(\th),$$ so that $\th f(\th)-F(\th)\geq \thl f(\th) \geq 0$ for all $\th\in (\thl, \thh)$. It follows that $h'(t)\geq 0$, and so $h(t)$ is increasing on $(0, 1)$. 

\item If $f(\th)$ is decreasing on $(\thl, \thh)$, then $$F(\th)=\int_{\thl}^\th f(t)dt \geq \int_{\thl}^\th f(\th)dt=(\th-\thl)f(\th),$$
so that $\th f(\th)-F(\th)\leq \thl f(\th)$ for all $\th\in (\thl, \thh)$. With the additional condition that $\thl=0$, we get that $h'(t)\leq 0$ and so $h(t)$ is decreasing on $(0,1)$.

\item[3,4.] If $\dfrac{f(\th)\th^2}{F^2(\th)}$ is increasing on $(\thl, \thh)$, then it must be that for all $\th\in (\thl, \thh)$, 
$$F^2(\th)(2\th f(\th)+\th^2f'(\th)) \geq f(\th)\th^22F(\th)f(\th).$$
But this is equivalent to $h''(t)\geq 0$. Analogously, $\frac{f(\th)\th^2}{F^2(\th)}$ being decreasing on $(\thl, \thh)$ is equivalent to $h''(t)<0$. 
\end{enumerate}
\end{proof}

Notice that if $F(.)$ is uniform on $\Th=[0,\thh]$, then $h(t)=\frac{1}{\thh}$ for all $t\in (0,1)$, in which case it satisfies all the above properties. Thus, $h(.)$ captures properties of the ability distribution relative to this benchmark uniform case, so that it is increasing if unproductive agents are more likely, decreasing if productive agents are more likely, concave if moderately productive agents are more likely, and convex if extreme (high or low) productivity agents are more likely.\\


Lastly, we note a key idea that we use repeatedly in our analysis. 
\begin{lemrep}
\label{lem:trick}
Suppose $f_1:[a,b]\to \mathbb{R}$ is such that there exists $c \in [a,b]$ so that $f_1(x)\leq 0$ for $x \leq c$ and $f_1(x) \geq 0$ for $x \geq c$. Then, for any increasing function $f_2:[a,b]\to \mathbb{R}$, $$\int_a^b f_1(x)f_2(x)dx \geq f_2(c)\int_a^b f_1(x)dx.$$
\end{lemrep}
\begin{proof}
Observe that
\begin{align*}
    \int_a^b f_1(x)f_2(x)dx&= \int_a^c f_1(x)f_2(x)dx+ \int_c^b f_1(x)f_2(x)dx\\
    &\geq \int_a^c f_1(x)f_2(c)dx+ \int_c^b f_1(x)f_2(c)dx\\
    &= f_2(c)\int_a^b f_1(x)dx
\end{align*}
\end{proof}

The claim clearly holds when $f_2(.)$ is a constant function. If $f_2(.)$ is increasing, it puts greater weight to the region of $f_1(.)$ where it is positive as compared to where $f_1(.)$ is negative, and therefore, the integral is greater in comparison to the case where the weight is uniform over the entire region.  
\section{Equilibrium}

In this section, we study the equilibrium effort induced by different contests, with a focus on understanding how manipulating different prizes and increasing competitiveness can influence expected equilibrium effort.\\

To begin, we note the symmetric Bayes-Nash equilibrium of the game induced by an arbitrary contest $v$. In a similar setup but with $\thl>0$,  \citet*{moldovanu2001optimal} characterized the equilibrium of the contest game, showing that more productive agents exert greater effort. The same characterization extends to this case, where $\thl$ might possibly be $0$.

\begin{lemrep}
\label{lem:equilibrium}
For any contest $v=\{v_1, \dots, v_{n}\}$, there is a unique symmetric Bayes-Nash equilibrium and it is such that for any $\th\in \Th$,
$$X_{v}(\th)=g\left(\sum_{i=1}^n v_im_i(\th)\right),$$
where
\begin{equation}
\label{eq:marginal_effects}
m_i(\th)=-\int_{F(\th)}^1 \dfrac{ p'_i(t)}{F^{-1}(t)}dt.
\end{equation}
\end{lemrep}
\begin{proof}
Suppose $n-1$ agents are playing a strategy $X: \Th \to \mathbb{R}_+$ so that if an agent's type is $\th$, it exerts effort $X(\th)$. Further suppose that $X(\th)$ is decreasing in $\th$. Now we want to find the remaining agent's best response to this strategy of the other agents. If the agent's type is $\th$ and it pretends to be an agent of type $\hat{\th} \in \Th$, its payoff is
$$\sum_{i=1}^n v_ip_i(F(\hat{\th}))-\th c(X(\hat{\th})).$$ 
Taking the first-order condition, we get
\begin{equation}
\label{eq:foc}
\sum_{i=1}^n v_ip_i'(F(\hat{\th}))f(\hat{\th})-\th c'(X(\hat{\th})) X'(\hat{\th})=0.
\end{equation}

Now we can plug in $\hat{\th}=\th$ to get the condition for $X(\th)$ to be a symmetric Bayes-Nash equilibrium. Doing so, we get 
$$\sum_{i=1}^n v_ip_i'(F(\th))f(\th)-\th c'(X(\th)) X'(\th)=0$$ so that
$$\int_{\th}^{\thh} c'(X(\hat{\th})) X'(\hat{\th}) d\hat{\th}=\int_\th^{\thh} \dfrac{\sum_{i=1}^n v_ip_i'(F(\hat{\th}))}{\hat{\th}}f(\hat{\th})d\hat{\th}.$$
Using the boundary condition $X(\thh)=0$, and $c(0)=0$, we obtain
\begin{align*}
    c(X(\th))&=-\int_\th^{\thh} \dfrac{\sum_{i=1}^n v_ip_i'(F(\hat{\th}))}{\hat{\th}}f(\hat{\th})d\hat{\th}\\
    &=-\int_{F(\th)}^{1} \dfrac{\sum_{i=1}^n v_ip_i'(t)}{F^{-1}(t)}dt &(\text{Substituting } F(\hat{\th})=t)\\
    &=\sum_{i=1}^n v_i \left[-\int_{F(\th)}^{1} \dfrac{p_i'(t)}{F^{-1}(t)}dt\right].
\end{align*}
This gives us a precise candidate equilibrium $X_v(\th)$, and we can verify that $X_v'(\th)<0$ for all $\th\in (\thl, \thh)$ so that $X_v(\th)$ is indeed decreasing in marginal cost. Now we can repeat the above exercise with this specific candidate strategy. By construction, the first-order condition in Equation \eqref{eq:foc} will be uniquely satisfied at $\hat{\th}=\th$ for any agent of type $\th\in (\thl, \thh)$. Further,  differentiating the left-hand side of Equation \eqref{eq:foc}, we get
$$\sum_{i=1}^n v_i\left[p_i'(F(\hat{\th}))f'(\hat{\th})+f^2(\hat{\th})p_i''(F(\hat{\th}))\right]-\th \left[c'(X_v(\hat{\th})) X_v''(\hat{\th})+c''(X_v(\hat{\th})) X_v'^2(\hat{\th})\right],$$
which is $c'(X_v(\th))X_v'(\th) <0$ when $\hat{\th}=\th$ so that the second-order condition is also satisfied. 
\end{proof}

Notice that by Leibniz rule, 
$$\dfrac{\partial X_v(\th)}{\partial \th} = g'\left(\sum_{i=1}^n v_im_i(\th)\right)\left(\sum_{i=1}^n v_ip'_i(F(\th))\right)\dfrac{f(\th)}{\th}.$$
Since $\sum_{i=1}^k p_i(t)$ is decreasing in $t$ for all $k \in \{1, \dots, n\}$, and $v_1\geq v_2\geq \dots \geq v_n$, we get that  $\sum_{i=1}^n v_ip'_i(F(\th))<0$. As $g=c^{-1}$ is strictly increasing, it follows that $X_v(\th)$ is indeed strictly decreasing in $\th$, so that more productive agents exert greater effort. \\

From the characterization, one can verify that for any sequence of distributions $F_1, F_2, \dots$, converging pointwise to a distribution $F$, the corresponding sequence of equilibria also converges to the equilibrium under $F$. This means that, for instance, the equilibrium under the uniform distribution on $\Th=[0,\thh]$ is the limit of sequence of equilibria under uniform distributions $F_k$ on $[\thl_k, \thh]$, where $\lim_{k \to \infty} \thl_k=0$. This provides a useful alternative interpretation for equilibria in cases where $\thl=0$. More precisely, the equilibrium in such a case can serve as a useful approximation to equilibrium in settings where $\thl$ is positive but small. When $\thl=0$, we will let $X_v(0)=\lim_{\th \to 0 } X_v(\th)$, noting that this limit might be infinite.\footnote{When $\thl=0$, notice that if $\lim_{\th \to 0 } X_v(\th)$ is finite, the agent of type $\th=0$ can choose any $x\geq \lim_{\th \to 0 } X_v(\th)$ in equilibrium. But since $\th=0$ is a measure zero event, the precise choice of $X_v(0)$ does not matter for our results, and we define it as above for concreteness.} \\

From this equilibrium characterization, the expected effort induced by any contest $v$ is
$$
\E[X_v(\th)]=\E\left[g\left(\sum_{i=1}^n v_im_i(\th)\right)\right].
$$

\subsection{Increasing prizes}

In this subsection, we study the effect of manipulating individual prizes on expected equilibrium effort. From above, the marginal effect of increasing value of any prize $i$ is
\begin{equation}
\label{eq:general_marginal_effect}  
\dfrac{\partial \E[X_v(\th)]}{\partial v_i}=\E\left[g'\left(\sum_{i=1}^n v_im_i(\th)\right) m_i(\th)\right].
\end{equation}

Now one might reasonably suspect that this effect varies across prizes, and in particular, expect that increasing the value of top ranked prizes should encourage effort, while increasing the value of bottom ranked prizes should discourage effort. This intuition indeed holds true in the extreme case where top ranked prize is the first prize and the bottom ranked prize is the last prize. To see why, observe that $p_1'(.)<0$ and $p_n'(.)>0$, and so it follows from Equation \eqref{eq:marginal_effects} that $m_1(.)\geq 0$ and $m_n(.)\leq 0$. Thus, increasing the first prize always encourages effort from all agents while increasing the last prize always discourages effort from all agents. In the remainder of this subsection, we focus on the effect of increasing intermediate prizes on effort, showing in particular that the effect actually depends on the underlying distribution of abilities.\\

We first focus on the case where cost is linear. In this case, it follows from Equation \eqref{eq:general_marginal_effect} that the marginal effect of increasing prize $i$ is simply $\E[m_i(\th)]$. The following result identifies conditions on distributions under which this effect may be positive, or negative.

\begin{thmrep}
\label{thm:linear_absolute_effects}
For any intermediate prize $i\in \{2, \dots, n-1\}$, the following hold:
\begin{enumerate}
\item If $h(t)$ is increasing on $(0, 1)$, then $\E[m_i(\theta)]\geq 0$.
\item If $h(t)$ is decreasing on $(0, 1)$, then $\E[m_i(\theta)]\leq 0$.
\end{enumerate}
\end{thmrep}
\begin{proof}
Consider first the case where $\thl>0$. In this case, for any prize $i\in \{1,\dots, n\}$, 
\begin{align*}
\E[m_i(\th)]&=\int_{\thl}^{\thh} m_i(\th)f(\th)d\th\\
&=m_i(\th)F(\th) |_{\thl}^{\thh}-\int_{\thl}^{\thh} m_i'(\th)F(\th)d\th &(\text{Integration by parts})\\
&=-\int_{\thl}^{\thh} m_i'(\th)F(\th)d\th &(m_i(\thh)=0, F(\thl)=0)\\
&=-\int_{\thl}^{\thh} \dfrac{p_i'(F(\th))f(\th)}{\th}F(\th)d\th &(\text{Leibniz rule})\\
&=-\int_0^1 \dfrac{p_i'(t)}{F^{-1}(t)}tdt &(\text{Substituting } F(\th)=t)\\
&=-\int_0^1 p_i'(t)h(t)dt.
\end{align*}

Now for any intermediate prize $i\in \{2, \dots, n-1\}$, observe that $\int_0^1 p_i'(t)dt=0$, and also, there exists $c \in [0,1]$ such that $p_i'(t)\geq 0$ for $t\leq c$ and $p_i'(t)\leq 0$ for $t \geq c$. 

\begin{enumerate}

\item If $h(t)$ is increasing on $(0,1)$, we can apply Lemma \ref{lem:trick} with $f_1(t)=-p_i'(t)$, and $f_2(t)=h(t)$ to get that $$\E[m_i(\th)]\geq h(c)\int_0^1 (-p_i'(t))dt= 0.$$

\item If $h(t)$ is decreasing on $(0,1)$, we can apply Lemma \ref{lem:trick} with $f_1(t)=-p_i'(t)$, and $f_2(t)=-h(t)$ to get that $$-\E[m_i(\th)]\geq -h(c)\int_0^1 (-p_i'(t))dt= 0.$$
\end{enumerate}
This proves the result for the case where $\thl>0$. \\

Now consider the case where $\thl=0$. In this case, we can again directly use the above argument, except that we need to additionally ensure that $\lim_{\theta \to 0} m_i(\theta)F(\theta) = 0$. This is because even though $F(\thl)=0$, it might be the case that $\lim_{\theta \to 0} m_i(\theta)=\infty$. For this case,
\begin{align*}
\lim_{\theta \to 0} m_i(\theta)F(\theta)&=\lim_{\theta \to 0} \dfrac{m_i(\theta)}{\frac{1}{F(\theta)}}\\
&=\lim_{\theta \to 0} -\dfrac{m_i'(\theta)F^2(\theta)}{f(\theta)} &(\text{L'Hospital's rule})\\
&=\lim_{\theta \to 0} -\dfrac{p_i'(F(\theta))F^2(\theta)}{\theta} &(\text{Leibniz rule})\\
&=\lim_{t \to 0} -\dfrac{p_i'(t)t^2}{F^{-1}(t)}  &(\text{Substituting } F(\th)=t)\\
&=0 &(\text{Assumption \ref{ass:geniusrare}})
\end{align*}
Thus, Assumption \ref{ass:geniusrare} ensures that in case $\thl=0$, $\lim_{\theta \to 0} m_i(\theta)F(\theta)=0$, and we can indeed apply the steps in the argument for $\thl>0$ to this case as well, thus proving the result.
\end{proof}

To prove Theorem \ref{thm:linear_absolute_effects}, we show that for any prize $i\in \{1, \dots, n\}$,
\begin{equation}
\label{eq:expect_marginal_effects}
\E[m_i(\th)]= -\int_0^1 p_i'(t)h(t)dt.
\end{equation}
From here, the single-crossing property of $p_i'(.)$ for any intermediate prize $i$ leads to its effect being distribution dependent, with Lemma \ref{lem:trick} enabling us to obtain the specific sufficient conditions in the result. Intuitively, unlike the first and last prize, increasing any intermediate prize has contrasting effects on different agent types. It encourages effort from unproductive agents while discouraging effort from the productive agents (as illustrated in Figure \ref{fig:eqbm}), and thus, the overall effect  depends on the relative likelihood of these agents. In particular, from the sufficient conditions in Lemma \ref{lem:conditions}, we get that if the density is increasing in marginal costs so that the distribution is predominantly unproductive, the overall effect is positive, and if the density is decreasing in marginal costs so that the distribution is predominantly productive (with the additional condition that extremely productive agents are possible), the overall effect is negative.\\

Now for the case of general costs, we show that the effect of increasing intermediate prize can sometimes be informed by the effect of such a transformation under linear costs. The idea is that with general costs, one can essentially reinterpret the Bayesian game as one in which the agents are directly choosing effort cost $c(x)$ instead of effort $x$. With this interpretation, $\E[m_i(\th)]$ captures not just the effect on expected effort under linear costs, but more generally, the effect on expected effort costs $\E[c(X_v(\th))]$. And as the following result shows, it is sometimes sufficient to recover the qualitative effect on expected effort as well.

\begin{proprep}
\label{prop:general_absolute_effects}
For any intermediate prize $i\in \{2, \dots, n-1\}$, the following hold:
\begin{enumerate}
    \item If $c(.)$ is convex and $\E[m_i(\th)]\geq 0$, then for any contest $v$, $\dfrac{\partial \E[X_v(\th)]}{\partial v_i}\geq 0$. 
    \item If $c(.)$ is concave and $\E[m_i(\th)]\leq 0$, then for any contest $v$, $\dfrac{\partial \E[X_v(\th)]}{\partial v_i}\leq 0$.
\end{enumerate}
\end{proprep}
\begin{proof}
From Equation \eqref{eq:general_marginal_effect}, we have that 
\begin{align*}
\dfrac{\partial \E[X_v(\th)]}{\partial v_i}&=\int_{\thl}^{\thh} g'\left(\sum_{i=1}^n v_im_i(\th)\right) m_i(\th) f(\th) d\th.
\end{align*}
Observe that $\int_{\thl}^{\thh} m_i(\th) f(\th) d\th=\E[m_i(\th)]$, and moreover, there exists $\hat{\th} \in [\thl,\thh]$ such that  $m_i(\th) f(\th)\leq 0$ for $\th \leq \hat{\th}$ and $m_i(\th) f(\th)\geq 0$ for $\th \geq \hat{\th}$. 
Now we can prove the claims in order.
\begin{enumerate}[leftmargin=*]
    \item Consider any convex cost function $c(.)$ and any contest $v$. It follows that $g=c^{-1}$ is concave, so that $g'(.)$ is an decreasing function, and thus, $g'\left(\sum_{i=1}^n v_im_i(\th)\right)$ is increasing in $\th$. Now we can apply Lemma \ref{lem:trick} with $f_1(\th)=m_i(\th)f(\th)$, and $f_2(\th)=g'\left(\sum_{i=1}^n v_im_i(\th)\right)$ to get that $$\dfrac{\partial \E[X_v(\th)]}{\partial v_i}\geq g'\left(\sum_{i=1}^n v_im_i(\hat{\th})\right)\E[m_i(\th)].$$
    In particular, we get that if $\E[m_i(\th)]\geq 0$, then $\dfrac{\partial \E[X_v(\th)]}{\partial v_i}\geq 0$.
    \item  Consider any concave cost function $c(.)$ and any contest $v$. It follows that $g=c^{-1}$ is convex, so that $g'(.)$ is an increasing function, and thus, $g'\left(\sum_{i=1}^n v_im_i(\th)\right)$ is decreasing in $\th$. Now we can apply Lemma \ref{lem:trick} with $f_1(\th)=m_i(\th)f(\th)$, and $f_2(\th)=-g'\left(\sum_{i=1}^n v_im_i(\th)\right)$ to get that $$-\dfrac{\partial \E[X_v(\th)]}{\partial v_i}\geq -g'\left(\sum_{i=1}^n v_im_i(\hat{\th})\right)\E[m_i(\th)].$$
    In particular, we get that if $\E[m_i(\th)]\leq 0$, then $\dfrac{\partial \E[X_v(\th)]}{\partial v_i}\leq 0$.
\end{enumerate}
\end{proof}

To prove Proposition \ref{prop:general_absolute_effects}, we use the single-crossing property of $m_i(\th)$, and then apply Lemma \ref{lem:trick} on the representation of expected marginal effects in Equation \eqref{eq:general_marginal_effect}. Intuitively, we know that increasing an intermediate prize $i$ encourages effort from unproductive agents, while discouraging effort from productive agents. Since unproductive agents exert lower effort than productive agents, if the change in expected effort cost is positive under convex costs, it must be that the effort of unproductive agents increases more than the decrease in the effort of the productive agents. An analogous reasoning applies when cost is concave and the expected effort costs decrease.\footnote{Our results for the effect of increasing intermediate prizes have implications for the design problem where the designer can costlessly award any number of homogeneous prizes. This problem was studied by \citet*{liu2017optimal} under linear costs, and we can use Proposition \ref{prop:general_absolute_effects} to extend the results to more general cost functions.}

\subsection{Increasing competition}

In this subsection, we study the effect of increasing competition on expected equilibrium effort. Following the recent literature in contest design (\citet*{fang2020turning}), we say a contest $v$ is more competitive than $w$ if $v$ can be obtained from $w$ by a sequence of transfers from worse ranked prizes to better ranked prizes, and we investigate the effect of such transfers on expected effort. From Equation \eqref{eq:general_marginal_effect}, the marginal effect of transferring value of from a worse-ranked prize $j \in \{1, \dots, n\}$ to a better-ranked prize $i\in \{1, \dots, n\}$ is
\begin{equation}
\label{eq:general_relative_effect}  
\dfrac{\partial \E[X_v(\th)]}{\partial v_i}-\dfrac{\partial \E[X_v(\th)]}{\partial v_j}=\E\left[g'\left(\sum_{i=1}^n v_im_i(\th)\right) \left(m_i(\th)-m_j(\th)\right)\right].
\end{equation}

We already know that increasing the first prize and decreasing the last prize encourages effort from all agents, and thus, increasing competition by transferring value from last prize to the first prize encourages effort. In the remainder of this subsection, we investigate the effect of increasing competition via transfers that involve intermediate prizes. \\

As before, we first focus on the case where the cost is linear. In this case, it follows from Equation \eqref{eq:general_relative_effect} that the effect of increasing competition is simply $\E[m_i(\th)-m_j(\th)]$. In a model with $\thl>0$, \citet{moldovanu2001optimal} showed that transferring value to the first prize always encourages effort (i.e.,  $\E[m_1(\th)]>\E[m_j(\th)]$ for any $j\in \{2, \dots, n\}$), thus establishing the optimality of the winner-takes-all contest for maximizing expected effort under linear costs. While this suggests that increasing competition might generally encourage effort, the following result shows that this isn't necessarily the case.

\begin{thmrep}
\label{thm:relative_effects}

For any pair of prizes $i, j \in \{2, \dots, n-1\}$ with $i \leq j$, the following hold:
\begin{enumerate}
\item $\E[m_1(\theta)]> \E[m_j(\theta)]$.
\item If $h(t)$ is concave on $(0,1)$, then  $\E[m_i(\theta)]\geq \E[m_j(\theta)]$.
\item If $h(t)$ is convex on $(0,1)$, then $\E[m_i(\theta)]\leq \E[m_j(\theta)]$.
\end{enumerate}
\end{thmrep}

\begin{figure}[!h]
    \centering
    \subfloat[\centering Increasing prizes]{{\includegraphics[width=6cm, height=5cm]{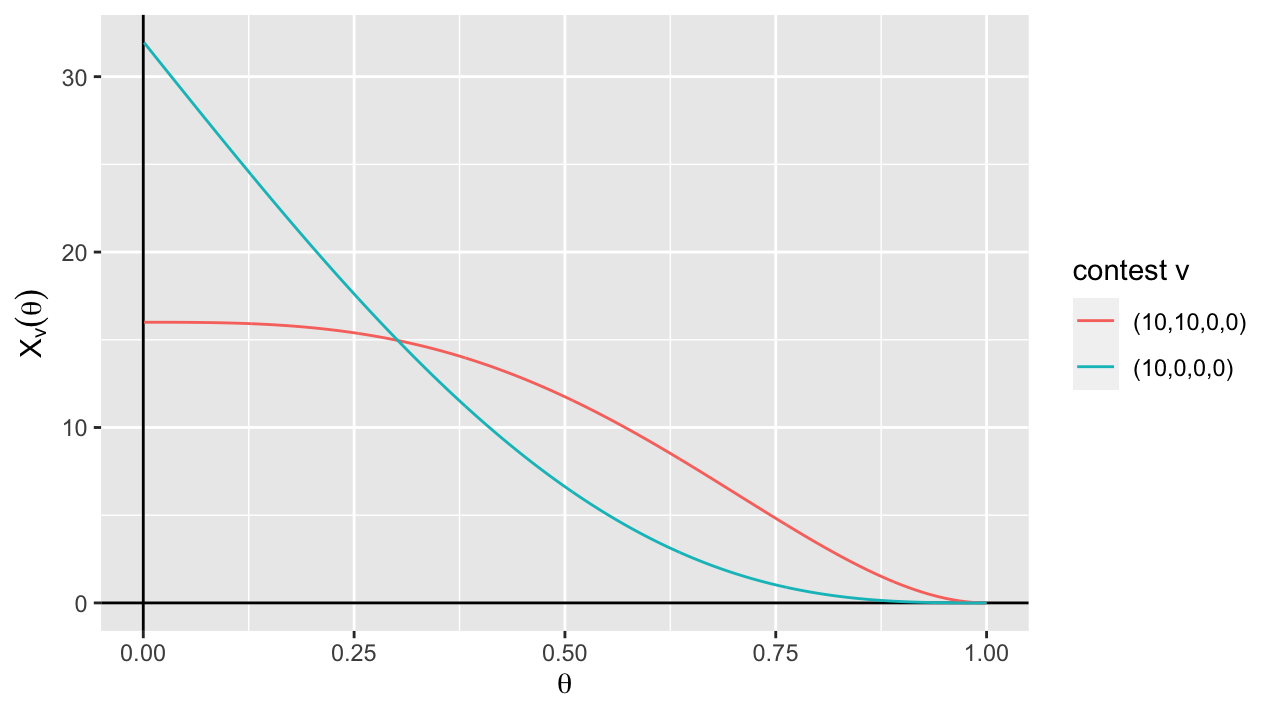} }}%
    \qquad
    \subfloat[\centering Increasing competition]{{\includegraphics[width=6cm, height=5cm]{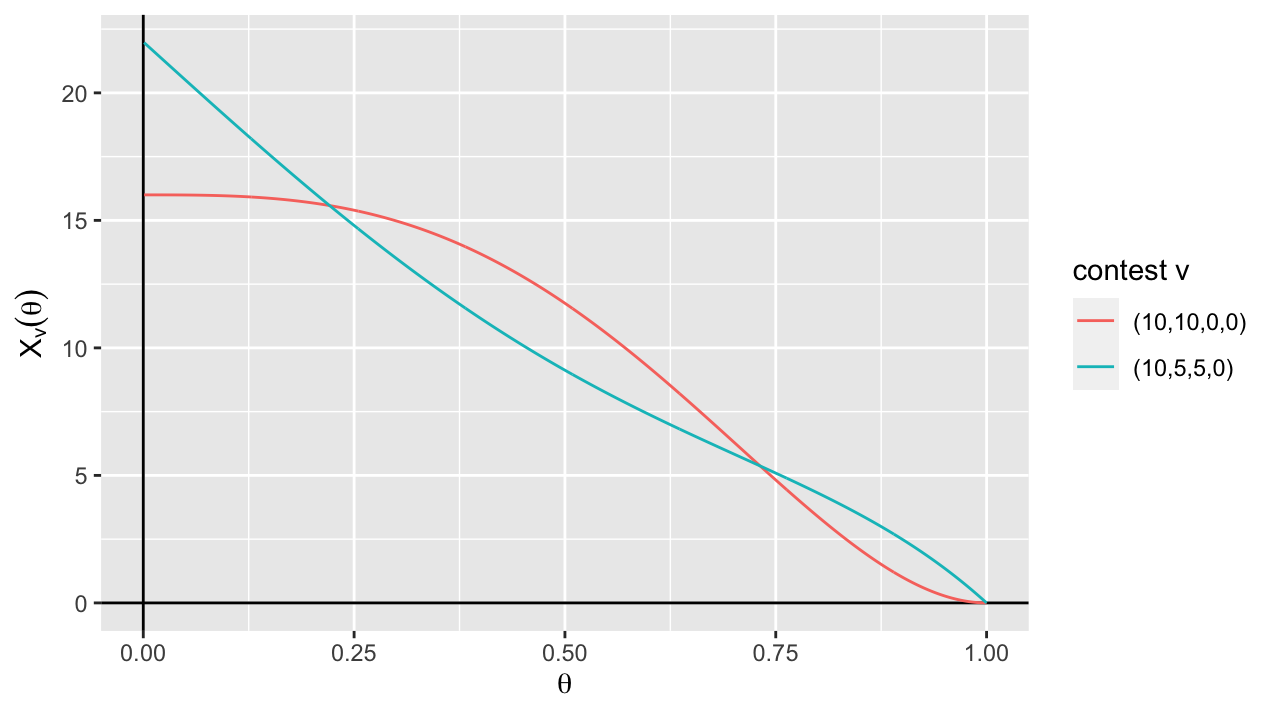} }}%
    \caption{Effect on manipulating prizes on effort with $n=4$, $\Th=[0,1]$, $F(\theta)=\theta^2$, $c(x)=x$.}%
    \label{fig:eqbm}%
\end{figure}

\begin{proof}
We prove the claims in order. 
\begin{enumerate}[leftmargin=*]
    \item For any $j\in \{2, \dots, n-1\}$, it follows from Equation \eqref{eq:expect_marginal_effects} that
    $$\E[m_1(\th)]-\E[m_j(\th)]=\int_0^1 \left(p_j'(t)-p_1'(t)\right) \frac{t}{F^{-1}(t)}dt.$$
    From Lemma \ref{lem:calculation},
    \begin{align*}
        \int_0^1 \left(p_j'(t)-p_1'(t)\right)tdt &= p_j(1)-{n-1 \choose j-1} \B(j, n-j+1)-p_1(1)+\B(1, n)\\
        &=0,
    \end{align*}
    and moreover, there exists $c \in [0,1]$ such that $ (p_j'(t)-p_1'(t))t\geq 0$ for $t\leq c$ and $ (p_j'(t)-p_1'(t))t\leq 0$ for $t\geq c$. Since $\frac{1}{F^{-1}(t)}$ is strictly decreasing in $t$ on $(0,1)$, we can apply Lemma \ref{lem:trick} with $f_2(t)=-\frac{1}{F^{-1}(t)}$ and $f_1(t)=-(p_j'(t)-p_1'(t))t$ to get that
    $$\E[m_1(\th)]-\E[m_j(\th)]> -\frac{1}{F^{-1}(c)}\int_0^1 -\left(p_j'(t)-p_1'(t)\right) tdt=0.$$

\item[2,3.] Consider first the case where $\thl>0$. For any $i, j \in \{2, \dots, n-1\}$ with $i<j$, it follows from Equation \eqref{eq:expect_marginal_effects} that
\begin{align*}
\E[m_i(\th)]-\E[m_j(\th)]&=\int_0^1 \left(p_j'(t)-p_i'(t)\right) h(t)dt\\
&=\left(p_j(t)-p_i(t)\right)h(t)|^1_0-\int_0^1 (p_j(t)-p_i(t))h'(t)dt \\
&=\int_0^1 (p_i(t)-p_j(t))h'(t)dt.
\end{align*}

Observe that $\int_0^1 (p_i(t)-p_j(t))dt=0$, and moreover, there exists $c \in [0,1]$ such that  $p_i(t)-p_j(t)\geq 0$ for $t\leq c$ and $p_i(t)-p_j(t)\leq 0$ for $t\geq c$. 

If $h(t)$ is concave on $(0,1)$, $h'(t)$ is monotone decreasing on $(0,1)$, and we can apply Lemma \ref{lem:trick} with $f_1(t)=-(p_i(t)-p_j(t))$, and $f_2(t)=-h'(t)$ to get that $$\E[m_i(\th)] - \E[m_j(\th)]\geq -h'(c)\int_0^1 -(p_i(t)-p_j(t))dt= 0.$$

If $h(t)$ is convex on $(0,1)$, $h'(t)$ is monotone increasing on $(0,1)$, and we can apply Lemma \ref{lem:trick} with $f_1(t)=-(p_i(t)-p_j(t))$, and $f_2(t)=h'(t)$ to get that $$-\left[\E[m_i(\th)] - \E[m_j(\th)]\right]\geq h'(c)\int_0^1 -(p_i(t)-p_j(t))dt= 0.$$

This proves the result for the case where $\thl>0$. \\

Now consider the case where $\thl=0$. In this case, we can again directly use the above argument, except that we need to additionally ensure that $\lim_{t \to 0} (p_j(t)-p_i(t))h(t) = 0$. This is because even though $p_j(0)-p_i(0)=0$, it might be the case that $\lim_{t \to 0} h(t)=\infty$. For this case, observe that for any $k\in \{2, \dots, n-1\}$,
\begin{align*}
    \lim_{t \to 0} p_k(t)h(t) &= \lim_{t \to 0} {n-1 \choose k-1}\frac{t^{k}(1-t)^{n-k}}{F^{-1}(t)}\\
    &=0 &(\text{Assumption \ref{ass:geniusrare}}),
\end{align*}
which implies that for any $i, j \in \{2, \dots, n-1\}$, 
$\lim_{t \to 0} (p_j(t)-p_i(t))h(t) = 0$. Thus, Assumption \ref{ass:geniusrare} ensures that even in the case that $\thl=0$, we can apply the steps in the argument for $\thl>0$, thus proving the result.
\end{enumerate}
\end{proof}     
To prove Theorem \ref{thm:relative_effects}, we show that for any pair $i,j \in \{2, \dots, n-1\}$ with $i<j$,  
\begin{equation}
\label{eq:expect_difference_marginal_effects}
\E[m_i(\th)]-\E[m_j\th)]=\int_0^1 (p_i(t)-p_j(t))h'(t)dt.
\end{equation}
From here, we use the single-crossing property of $p_i(t)-p_j(t)$ which, together with Lemma \ref{lem:trick}, leads to the sufficient conditions in the result.\footnote{While Theorem \ref{thm:relative_effects} does not say anything about the effect of transferring value from the last prize to an intermediate prize, we illustrate through an example that even this transformation may discourage effort. Consider $\Th=[0,1]$ with $F(\theta)=\theta^p$ where $p>\frac{1}{2}$. For $n=3$ agents, we can use Lemma \ref{lem:calculation} to show that 
$\mathbb{E}[m_2(\theta)]=\frac{2p(p-1)}{(3p-1)(2p-1)}\text{ and } \mathbb{E}[m_3(\theta)]=\frac{-2p}{3p-1},$
so that $\mathbb{E}[m_2(\theta)]<\mathbb{E}[m_3(\theta)]$ if $\frac{1}{2}<p<\frac{2}{3}$.} Thus, even though increasing competition by transferring value from an intermediate prize to the first prize always encourages effort, there exist distributions for which transferring value to any other better-ranked intermediate prize (including the second prize) might actually discourage effort. Intuitively, such a transformation discourages effort from the least productive agents, while encouraging effort from the moderately productive agents (as illustrated in Figure \ref{fig:eqbm}). The effect on the most productive agents may be positive or negative. The overall effect of increasing competition is, therefore, determined by the likelihood of the moderately productive agents.\\

Now for the case of general costs, we again show that the effect of increasing competition can sometimes be informed by the effect of such a transformation under linear costs. As noted above, increasing competition can have a positive or negative effect on the effort of the most productive agents. In case the transformation encourages effort from the most productive agents, captured by the condition that the effort of the most productive agent increases, the effect under general costs may be informed by the effect under linear costs.\footnote{In recent work, \citet*{baranski2024contest} study the effect of competition in contests with finite type-spaces, and show that the effect under general costs may be informed by the effect under linear costs if the equilibrium utility of the most productive agent decreases. This is equivalent to the condition that the equilibrium effort of the most productive agent increases in our setting, since this agent always wins the first prize.}  
\begin{proprep}
\label{prop:general_relative_effects}
Consider any pair of prizes $i,j \in \{1, \dots, n-1\}$ with $i<j$. If $i=1$ or if $m_i(\thl)\geq m_j(\thl)$, the following hold:
\begin{enumerate}
    \item If $c$ is concave and $\E[m_i(\th)]\geq \E[m_j(\th)]$, then for any contest $v$, $\dfrac{\partial \E[X_v(\th)]}{\partial v_i}\geq \dfrac{\partial \E[X_v(\th)]}{\partial v_j}$. 
    \item If $c$ is convex and $\E[m_i(\th)]\leq \E[m_j(\th)]$, then for any contest $v$, $\dfrac{\partial \E[X_v(\th)]}{\partial v_i}\leq \dfrac{\partial \E[X_v(\th)]}{\partial v_j}$. 
\end{enumerate}
\end{proprep}
\begin{proof}
From Equation \eqref{eq:general_relative_effect}, we have that
\begin{align*}
\dfrac{\partial \E[X_v(\th)]}{\partial v_i} - \dfrac{\partial \E[X_v(\th)]}{\partial v_j} &=\int_{\thl}^{\thh} g'\left(\sum_{i=1}^n v_im_i(\th)\right) \left[m_i(\th) - m_j(\th)\right]f(\th) d\th.
\end{align*}

From Equation \eqref{eq:marginal_effects}, the effect on effort exerted by agent of type $\th\in \Th$ is captured by
\begin{align*}
    m_i(\th) - m_j(\th) &=\int_{F(\th)}^1 \dfrac{ p'_j(t) - p_i'(t)}{F^{-1}(t)}dt.
\end{align*}

From here, one can verify that
\begin{enumerate}
    \item $m_i(\thh)-m_j(\thh)=0$,
    \item $m_i(\thl)-m_j(\thl) = \int_{0}^{1} \frac{p'_j(t) - p_i'(t)}{F^{-1}(t)}dt$,
    \item $m_i'(\th)-m_j'(\th) = \left[p_i'(F(\th))-p_j'(F(\th))\right]\frac{f(\th)}{\th}$. 
\end{enumerate}

Observe that if $i=1$, $p_1'(t)-p_j'(t)$ is initially negative, and then positive. Thus, there exists a $\hat{\th}\in [\thl, \thh]$ such that $m_1(\th)-m_j(\th)\geq 0$ for $\th\leq \hat{\th}$, and $m_i(\th)-m_j(\th)\leq 0$ for $\th\geq \hat{\th}$.

And if $1<i<j<n$, then  $p_i'(t)-p_j'(t)$ is initially positive, then negative, and then again positive. Since we assume $m_i(\thl)-m_j(\thl)\geq 0$, these properties imply that there exists a $\hat{\th}\in [\thl, \thh]$ such that $m_i(\th)-m_j(\th)\geq 0$ for $\th\leq \hat{\th}$, and $m_i(\th)-m_j(\th)\leq 0$ for $\th\geq \hat{\th}$. Now we can prove the claims in order.
\begin{enumerate}[leftmargin=*]
    \item  Consider any concave cost function $c(.)$ and any contest $v$. It follows that $g=c^{-1}$ is convex, so that $g'(.)$ is an increasing function, and thus, $g'\left(\sum_{i=1}^n v_im_i(\th)\right)$ is decreasing in $\th$. Now we can apply Lemma \ref{lem:trick} with $f_1(\th)= - (m_i(\th)-m_j(\th))f(\th)$, and $f_2(\th)=-g'\left(\sum_{i=1}^n v_im_i(\th)\right)$ to get that $$\dfrac{\partial \E[X_v(\th)]}{\partial v_i} - \dfrac{\partial \E[X_v(\th)]}{\partial v_j}\geq -g'\left(\sum_{i=1}^n v_im_i(\hat{\th})\right)\left(-\left[\E[m_i(\th)-m_j(\th)]\right]\right).$$
    In particular, if $\E[m_i(\th)]\geq \E[m_j(\th)]$, then $\dfrac{\partial \E[X_v(\th)]}{\partial v_i} \geq \dfrac{\partial \E[X_v(\th)]}{\partial v_j}.$
    \item Consider any convex cost function $c(.)$ and any contest $v$. It follows that $g=c^{-1}$ is concave, so that $g'(.)$ is an decreasing function, and thus, $g'\left(\sum_{i=1}^n v_im_i(\th)\right)$ is increasing in $\th$. Now we can apply Lemma \ref{lem:trick} with $f_1(\th)= - (m_i(\th)-m_j(\th))f(\th)$, and $f_2(\th)=g'\left(\sum_{i=1}^n v_im_i(\th)\right)$ to get that $$ - \left[\dfrac{\partial \E[X_v(\th)]}{\partial v_i} - \dfrac{\partial \E[X_v(\th)]}{\partial v_j} \right]\geq g'\left(\sum_{i=1}^n v_im_i(\hat{\th})\right)\left(-\left[\E[m_i(\th)-m_j(\th)]\right]\right).$$
    In particular, if $\E[m_j(\th)]\geq \E[m_i(\th)]$, then $\dfrac{\partial \E[X_v(\th)]}{\partial v_j} \geq \dfrac{\partial \E[X_v(\th)]}{\partial v_i}.$
\end{enumerate}
    
\end{proof}

To prove Proposition \ref{prop:general_relative_effects}, we show that if the effort of the most productive agent increases, $m_i(\th)-m_j(\th)$ exhibits the single-crossing property. And then, we apply Lemma \ref{lem:trick} on the representation in Equation \eqref{eq:general_relative_effect}. Despite being somewhat limited in scope, Proposition \ref{prop:general_relative_effects} provides a convenient method to check if increasing competition would encourage or discourage effort in fairly general environments, and in particular, has implications for the classical design problem of allocating a fixed budget. For concave costs, since $\E[m_1(\th)]>\E[m_j(\th)]$ (Theorem \ref{thm:relative_effects}), it allows us to recover the optimality of the winner-takes-all contest (\citet*{moldovanu2001optimal}). And perhaps more interestingly, for convex costs, it reveals conditions under which the designer would optimally allocate equal value (if any) to each of the $n-2$ intermediate prizes.\footnote{We can provide examples where increasing competition encourages effort from the most productive agent ($m_i(\thl)\geq m_j(\thl)$) so that Proposition \ref{prop:general_relative_effects} applies. But it appears difficult to simultaneously have that these transformations discourage expected effort costs ($\E[m_i(\th)]\leq \E[m_j(\th)]$). If they exist, such examples would provide instances where allocating equal value to $n-2$ intermediate prizes is optimal, in contrast to previous literature which has generally identified conditions under which either the winner-takes-all contest or contests with multiple descending prizes are optimal (\citet*{moldovanu2001optimal,zhang2024optimal,olszewski2020performance}).}\\

Finally, in Theorems \ref{thm:linear_absolute_effects} and \ref{thm:relative_effects}, we observe that the sufficient conditions under which increasing intermediate prizes or competition encourages effort ($h(t)$ is increasing or concave) hold quite generally, encompassing cases like the the uniform distribution on $[\thl, \thh]$. However, the conditions under which they discourage effort ($h(t)$ is decreasing or convex) are more restrictive, as they imply the possibility of extremely productive agents $(\thl=0)$. While $\thl=0$ allows for a clean presentation of the results,  we note that the insights revealed about how the effect of these transformations depend on the underlying distribution hold more generally. We illustrate our results in Table \ref{table:examples} for parametric distributions $F(\th)=\th^p$ and $F(\th)=1-(1-\th)^q$ on $\Th=[0,1]$. But we can compute the expected marginal effects for the appropriately scaled and translated versions of these distributions on $\Th=[\thl, \thh]$, where $\thl$ is small but positive, and show that increasing some intermediate prizes would still discourage effort when productive agents are more likely than unproductive agents, and increasing competition by transferring value from some worse-ranked intermediate prizes to some better-ranked intermediate prizes would still discourage effort when moderately productive agents are unlikely.
 
\section{Grading schemes}

In this section, we discuss the application of our results to the design of grading schemes, interpreting grades as signals about the private abilities of the participating agents. 

Formally, the designer chooses a \textit{grading scheme}, defined by a sequence of strictly increasing natural numbers $g=(n_1, \dots, n_B)$ where $n_B=n$, with $n_0=0$ whenever needed. Given $g$, all agents simultaneously choose their effort. The agents are ranked according to their effort and awarded the corresponding grades (with ties broken uniformly at random), so that the top $n_1$ agents receive a common grade $s_1$, the next $n_2-n_1$ agents receive a common grade $s_2$, and so on, until the last $n_B-n_{B-1}$ agents who receive a common grade $s_B$.\\

The value of a grade for an agent is derived from the information it reveals about their private type.\footnote{Note that the informative value of a grade depends on the strategy of the agents. For instance, a trivial equilibrium strategy is one where none of the agents exert any effort. To see why this is an equilibrium, observe that under this strategy profile, the grade obtained by an agent does not reveal any information about its type, and thus, there is no incentive for any agent to exert higher effort and obtain a better grade. For our analysis, we focus on the case where the market interprets grades in a way that is consistent with the assumption that more productive agents exert greater effort.} In particular, if an agent obtains grade $s_b$, the market interprets this grade as a signal that this agent's private type $\th \in [\thl, \thh]$ ranks between $n_{b-1}+1$ and $n_b$ (both inclusive) in a random sample of size $n$. Together with the common prior $F:[\thl, \thh]\to [0,1]$, this signal induces a posterior on the agent's type and we assume that there is a strictly decreasing wage function so that the value of the grade equals the agent's expected wage under this posterior.
\begin{assumption}
\label{ass:expected_wage}
There is a strictly decreasing wage function $w:[\thl, \thh] \to \R_+$ such that under any grading scheme $g=(n_1, \dots, n_B)$, an agent's value from receiving grade $s_b$ is given by
$$\E[w(\th)|\th \in \{\th_{(n_{b-1}+1)}, \th_{(n_{b-1}+2)}, \dots, \th_{(n_b)}\}],$$ where $\th_{(i)}$ denotes the $i$th order statistic in an i.i.d. sample $\th_1,\th_2, \dots, \th_n$. 
\end{assumption}
Thus, under the grading scheme $g=(n_1, \dots, n_B)$, if agent $j$ chooses effort $x_j$ and receives grade $s_b$, its payoff is simply
$$\E[w(\th)|\th \in \{\th_{(n_{b-1}+1)}, \th_{(n_{b-1}+2)}, \dots, \th_{(n_b)}\}]-\th_jc(x_j).$$

We are interested in finding grading schemes that maximize expected equilibrium effort.

\subsection{Contests induced by grading schemes}

Under Assumption \ref{ass:expected_wage}, any grading scheme $g=(n_1, \dots, n_B)$ induces a unique and well-defined contest $v(g)=(v_1(g), \dots, v_n(g))$ among the $n$ agents, and we now obtain a useful representation for this contest. Observe that if a grading scheme identifies exactly an agent who is ranked $i\in \{1, \dots, n\}$, the value of this signal for the agents is 
\begin{equation}
v_i^*=\E[w(\th)|\th=\th_{(i)}].    
\end{equation}
It follows then that the rank-revealing grading scheme $g^{rank}=(1, \dots, n)$ induces a contest such that for any prize $i$,  $v_i(g^{rank})=v_i^*$. At the other extreme, the no information grading scheme $g^{no}=(n)$ induces a contest such that for any prize $i$, $v_i(g^{no})=\frac{\sum_{i=1}^n v_i^*}{n}$.
In general, as the following result shows, any arbitrary grading scheme $g$ induces a contest that has a useful representation in terms of $(v_1^*, v_2^*, \dots, v_n^*)$.

\begin{lemrep}
\label{lem:grades_to_prizes}
A grading scheme $g=\left(n_1, \dots, n_B\right)$ induces a contest $v(g)$ such that for any prize $i\in \{1, \dots, n\}$,
$$
v_i(g)=\frac{v_{n_{b-1}+1}^*+v_{n_{b-1}+2}^*+\cdots+v_{n_b}^*}{n_b-n_{b-1}}
$$
where $b \in \{1, \dots, B\}$ is such that $n_{b-1}<i \leq n_b$.
\end{lemrep}
\begin{proof}
By Assumption \ref{ass:expected_wage}, if a grading scheme identifies exactly an agent who is ranked $i\in \{1, \dots, n\}$, the value of prize $i$ under this grading scheme is $$v_i^*=\E[w(\th)|\th=\th_{(i)}].$$

Notice that for any pair of prizes $i,j$ with $i<j$, $\th_{(j)}$ stochastically dominates $\th_{(i)}$, and since the wage function $w:[\thl, \thh]\to \R_+$ is strictly decreasing, it must be that $v_i^*>v_j^*$. Now consider any arbitrary grading scheme $g=(n_1, \dots, n_B)$. If an agent is ranked $i\in \{1, \dots, n\}$, it will receive a grade $s_b$ where $b$ must be such that $n_{b-1}<i<n_b$. Then, the market learns that the agent's rank must be one of $\left\{n_{b-1}+1, \ldots, n_b\right\}$, and moreover, it is equally likely to be ranked at any of these $n_b-n_{b-1}$ positions. Since the value of a grade is equal to the agent's expected wage under the posterior induced by the grade, the result follows.
\end{proof}
From Lemma \ref{lem:grades_to_prizes}, it is easy to see that if grading scheme $g'$ is more informative than grading scheme $g$, captured by the fact that $g$ is a subsequence of $g'$, the contest $v(g')$ is more competitive than the contest $v(g)$. \\

Moreover, observe that under any grading scheme $g=(n_1, \dots, n_B)$, the sum of all prizes in the contest induced by $g$ is $$\sum_{i=1}^n v_i(g)=\sum_{i=1}^n v_i^*=n \E[w(\th)].$$
Thus, as in the classical contest design problem, the designer essentially has a fixed budget of $v_1^*+v_2^*+\dots+v_n^*$ that it can distribute across $n$ prizes. However, unlike the classical problem, the designer cannot arbitrarily distribute this budget and is constrained to choose between the finitely many distributions that can be induced via grading schemes.

\subsection{Effort-maximizing grading schemes}

In this subsection, we will study effort-maximizing grading schemes. We will focus mostly on the design problem under linear costs, and briefly discuss implications of our results for general costs.\\

In the classical case, where the designer can allocate the entire budget arbitrarily, the optimal distribution under linear costs entails allocating the entire budget to the first prize, irrespective of the distribution of abilities. However, Lemma \ref{lem:grades_to_prizes} implies certain bounds on the values that can be allocated to different prizes, and in particular, constrains the value of the first prize to be at most $v_1^*$. While an effort-maximizing grading rule will always have the property that the first prize is allocated its maximum possible value of $v_1^*$, meaning that the best-performing agent is always uniquely identified, the optimal distribution of the remaining budget of $v_2^*+\dots+v_n^*$ among the remaining $n-1$ prizes depends on how manipulating these prizes and transferring value between these prizes effects effort. As we derived in Theorems \ref{thm:linear_absolute_effects} and \ref{thm:relative_effects}, these effects depend on the distribution of abilities. Consequently, the structure of the effort-maximizing grading scheme also depends in an important way on the underlying distribution of abilities.

\begin{thmrep}
\label{thm:grading}
Suppose the cost function is $c(x)=x$. 
\begin{enumerate}
    \item If $h(t)$ is concave, $\exists k$ such that $g=(1,2, \dots, k, n)$  is optimal. In case $h(t)$ is increasing as well, $g=(1,2, \dots, n)$ is optimal.
    \item If $h(t)$ is convex, $\exists k$ such that $g=(1,k, n)$ is optimal.  In case $h(t)$ is increasing as well, $g=(1,n-1, n)$ is optimal.
\end{enumerate}
\end{thmrep}
\begin{proof}
Let $\l_i=\mathbb{E}[m_i(\th)]$ from Equation \eqref{eq:expect_marginal_effects} and let $v_i^*=\mathbb{E}[w(\th)|\th=\th_{(i)}]$. Then, by Lemma \ref{lem:grades_to_prizes},  the expected effort induced by an arbitrary grading scheme $g=(n_1, \dots, n_B)$ is 
$$\sum_{i=1}^n \l_i v_i(g),$$
where $$v_i(g)=\frac{v_{n_{b-1}+1}^*+v_{n_{b-1}+2}^*+\cdots+v^*_{n_b}}{n_b-n_{b-1}}
$$
and $b$ is such that $n_{b-1}<i \leq n_b$. Let $g=(n_1, n_2, \dots, n_B)$ with $n_B=n$ denote the effort-maximizing grading scheme. \\

First, we show that $n_1=1$. Suppose towards a contradiction that $n_1>1$. Consider the grading scheme $g' = (1, n_1, n_2, \dots, n_B)$ which is the same as $g$ except that it uniquely identifies the best performing agent and pools the next $n_1-1$ agents together. Then, it follows from Lemma \ref{lem:grades_to_prizes} that $v(g')$ can be obtained from $v(g)$ by a sequence of transfers from intermediate prizes $2, 3, \dots, n_1$ to the first prize. And since $\l_1>\l_j$ for any $j \in \{2, \dots, n\}$ (Theorem \ref{thm:relative_effects}), each of these transfers leads to an increase in expected effort. Thus, the expected effort under $g'$ must be higher than $g$, which is a contradiction. Therefore, it must be that $n_1=1$, so that the best performing agent is uniquely identified. The full structure of $g$ depends on $F(.)$, and we now prove the claims in the result.

\begin{enumerate}[leftmargin=*]
\item  If $h(t)$ is concave, we know from Theorem \ref{thm:relative_effects} that $$\l_1>\l_2\geq \l_3\geq \dots \geq \l_{n-1}.$$

We want to show $\exists k$ such that $g=(1, 2, \dots, k, n)$, which means that agents who are not pooled together with the worst-performing agent must be uniquely identified. Consider an arbitrary grading scheme $g'=(1, n_2, n_3, \dots, n_{B-1}, n)$  and define $g''=(1, 2, 3, \dots, k, n)$ where $k=n_{B-1}$. Then, if $g'' \neq g'$,   $v(g'')$ can be obtained from $v(g')$ by a sequence of transfers from lower ranked intermediate prizes to better ranked intermediate prizes, each of which encourages effort. Thus, $g''$ induces greater effort than $g'$. It follows that the optimal grading scheme $g$ must take the form $(1,2, \dots, k, n)$. If $h(t)$ is increasing as well, we further have that $\l_{n-1}\geq 0 \geq \l_n$ (Theorem \ref{thm:linear_absolute_effects}), which leads to the rank-revealing scheme $g=(1, 2, \dots, n)$ being optimal.

\item If $h(t)$ is convex, we know from Theorem \ref{thm:relative_effects} that $$\l_1>\l_{n-1}\geq \l_{n-2}\geq \dots \geq \l_{2}.$$

We want to show that $\exists k$ such that $g=(1,k, n)$, which means that all agents, except the first, who are not pooled together with the worst-performing agent are pooled together with a common grade.  Consider an arbitrary grading scheme $g'=(1, n_2, n_3, \dots, n_{B-1}, n)$  and define $g''=(1, k, n)$ where $k=n_{B-1}$. Then, if $g'' \neq g'$,  $v(g')$ can be obtained from $v(g'')$ by a sequence of transfers from lower ranked intermediate prizes to better ranked intermediate prizes, each of which discourages effort. Thus, $g''$ induces greater effort than $g'$. It follows that the optimal grading scheme $g$ must take the form $(1, k, n)$. If $h(t)$ is increasing as well, we further have that $\l_{2}\geq 0 \geq \l_n$ (Theorem \ref{thm:linear_absolute_effects}), which leads to the grading scheme $g=(1, n-1, n)$ being optimal.

\end{enumerate}

\end{proof}

To prove Theorem \ref{thm:grading}, we use the fact that more informative grading schemes induce more competitive contests, and then exploit the effects of increasing competition from Theorem \ref{thm:relative_effects} to solve for the effort-maximizing grading schemes. More precisely, when $h(t)$ is concave so that moderately productive agents are likely, it follows that more informative schemes encourage effort. Thus,  the effort-maximizing grading scheme generally provides precise signals about the ranks of the agents, revealing exactly the rank of the top $k$ agents while pooling the remaining $n-k$ agents together with a common grade. In comparison, when $h(t)$ is convex so that moderately productive agents are unlikely, it follows that informative schemes discourage effort. Thus, the effort-maximizing grading scheme provides limited information about the ranks of the agents, pooling them by awarding at most three different grades. In case $h(t)$ is increasing, we obtain a more precise description with the effort maximizing grading scheme revealing all ranks exactly when $h(t)$ is concave, and pooling all intermediate ranks together with a common grade when $h(t)$ is convex. \\

Table \ref{table:examples} illustrates the results for two parametric classes of distributions on $\Th=[0,1]$, $F(\th)=\th^p$ with $p>\frac{1}{2}$, and $F(\th)=1-(1-\th)^q$ for $q>0$. Notice that for $p>1$, the density is bounded above by $p$, whereas if $\frac{1}{2}<p<1$, extremely productive agents become highly likely. Similarly, for $q>1$, the density is bounded above by $q$, but for $0<q<1$, extremely unproductive agents become highly likely. The bounded densities ensure that moderately productive are relatively likely, in which case the effort-maximizing grading schemes are highly informative. In comparison, when extreme productivity agents (high or low) become highly likely, we get that effort-maximizing grading schemes exhibit a coarse structure.\\

\begin{table}[h]
\begin{center}
\def\arraystretch{1.5}
\begin{tabular}{|c|c| c| c|} 
 \hline
 Density $f(\th)$  & $h(t)=\frac{t}{F^{-1}(t)}$ & Order ($\l_i=\E[m_i(\th)]$) & Grading scheme\\ 
 \hline
 \hline
 $p\th^{p-1}$, $p>1$ & incr., concave & $\l_1>\l_2>\dots >\l_{n-1}>0>\l_n$ & (1, 2, 3, \dots, n)\\ 
\hline
 $q(1-\th)^{q-1}$, $q>1$  & decr., concave &  $\l_1>0>\l_{2}>\dots> \l_{n-1}$; $0>\l_n$&  (1,2,\dots, k, n)\\ 
 \hline
 $q(1-\th)^{q-1}$, $0<q<1$ & incr., convex & $\l_1>\l_{n-1}>\dots>\l_2>0>\l_n$& (1, n-1, n) \\ 

 \hline
 $p\th^{p-1}$, $\frac{1}{2}<p<1$ & decr., convex & $\l_1>0>\l_{n-1}>\dots> \l_2$; $0>\l_n$ & (1,k,n)\\

 \hline
\end{tabular}
\caption{\label{parametric}Optimal grading schemes with linear costs for different distributions on $\Th=[0,1]$}
\label{table:examples}
\end{center}
\end{table}

Lastly, we discuss the implications of our results for the design of grading schemes under general costs. Focusing on environments where increasing competition encourages effort from the most productive agent (as required in Proposition \ref{prop:general_relative_effects}), we can solve for the effect of increasing competition on expected effort under linear costs (Equation \eqref{eq:expect_difference_marginal_effects}). If this effect is positive, more informative grading schemes would encourage effort under concave costs, and if this effect is negative, more informative grading schemes would discourage effort under convex costs (Proposition \ref{prop:general_relative_effects}).


\section{Conclusion}


We study how grading schemes influence effort in a model where grades provide signals about the private abilities of the participating agents. Towards this goal, we first investigate how manipulating individual prizes and increasing competition influences effort in contests with private abilities. Under linear costs, we identify ability distributions under which increasing prizes or increasing competition may encourage or discourage effort, and further identify conditions under which these effects persist even under more general cost functions. Our results suggest that the effect of increasing prizes is determined by the relative likelihood of productive and unproductive agents, while the effect of increasing competition is determined by the likelihood of moderately productive agents.\\

We then discuss implications for the design of grading schemes. Under our interpretation of grading schemes as information disclosure policies, we find that more informative grading schemes induce more competitive contests. Using our results on the effect of competition, we get that more informative grading schemes encourage effort when moderately productive agents are likely, and discourage effort when they are unlikely. Consequently, the effort-maximizing grading rule may range from a fully informative rank-revealing grading scheme that reveals exactly the rank obtained by each agent with a unique grade, to a coarse grading scheme that pools the performances of many agents together by awarding at most three different grades. 


\newpage
\bibliographystyle{ecta}

\bibliography{refs}

\begin{thebibliography}{}

\end{thebibliography}


\begin{thebibliography}{34}
\newcommand{\enquote}[1]{``#1''}
\expandafter\ifx\csname natexlab\endcsname\relax\def\natexlab#1{#1}\fi

\bibitem[\protect\citeauthoryear{Baranski and Goel}{Baranski and
  Goel}{2024}]{baranski2024contest}
\textsc{Baranski, A. and S.~Goel} (2024): \enquote{Contest design with a finite
  type-space: A unifying approach,} \emph{arXiv preprint arXiv:2410.04970}.

\bibitem[\protect\citeauthoryear{Barut and Kovenock}{Barut and
  Kovenock}{1998}]{barut1998symmetric}
\textsc{Barut, Y. and D.~Kovenock} (1998): \enquote{The symmetric multiple
  prize all-pay auction with complete information,} \emph{European Journal of
  Political Economy}, 14, 627--644.

\bibitem[\protect\citeauthoryear{Bevi{\'a} and Corch{\'o}n}{Bevi{\'a} and
  Corch{\'o}n}{2024}]{bevia2024contests}
\textsc{Bevi{\'a}, C. and L.~Corch{\'o}n} (2024): \emph{Contests: Theory and
  Applications}, Cambridge University Press.

\bibitem[\protect\citeauthoryear{Boleslavsky and Cotton}{Boleslavsky and
  Cotton}{2015}]{boleslavsky2015grading}
\textsc{Boleslavsky, R. and C.~Cotton} (2015): \enquote{Grading standards and
  education quality,} \emph{American Economic Journal: Microeconomics}, 7,
  248--279.

\bibitem[\protect\citeauthoryear{Brownback}{Brownback}{2018}]{brownback2018classroom}
\textsc{Brownback, A.} (2018): \enquote{A classroom experiment on effort
  allocation under relative grading,} \emph{Economics of Education Review}, 62,
  113--128.

\bibitem[\protect\citeauthoryear{Butcher, McEwan, and Weerapana}{Butcher
  et~al.}{2023}]{butcher2023making}
\textsc{Butcher, K., P.~J. McEwan, and A.~Weerapana} (2023): \enquote{Making
  the (letter) grade: The incentive effects of mandatory pass/fail courses,}
  \emph{Education Finance and Policy}, 1--24.

\bibitem[\protect\citeauthoryear{Chan, Hao, and Suen}{Chan
  et~al.}{2007}]{chan2007signaling}
\textsc{Chan, W., L.~Hao, and W.~Suen} (2007): \enquote{A signaling theory of
  grade inflation,} \emph{International Economic Review}, 48, 1065--1090.

\bibitem[\protect\citeauthoryear{Chowdhury, Esteve-Gonz{\'a}lez, and
  Mukherjee}{Chowdhury et~al.}{2023}]{chowdhury2023heterogeneity}
\textsc{Chowdhury, S.~M., P.~Esteve-Gonz{\'a}lez, and A.~Mukherjee} (2023):
  \enquote{Heterogeneity, leveling the playing field, and affirmative action in
  contests,} \emph{Southern Economic Journal}, 89, 924--974.

\bibitem[\protect\citeauthoryear{Corch{\'o}n}{Corch{\'o}n}{2007}]{corchon2007theory}
\textsc{Corch{\'o}n, L.~C.} (2007): \enquote{The theory of contests: a survey,}
  \emph{Review of economic design}, 11, 69--100.

\bibitem[\protect\citeauthoryear{Dubey and Geanakoplos}{Dubey and
  Geanakoplos}{2010}]{dubey2010grading}
\textsc{Dubey, P. and J.~Geanakoplos} (2010): \enquote{Grading exams: 100, 99,
  98,… or a, b, c?} \emph{Games and Economic Behavior}, 69, 72--94.

\bibitem[\protect\citeauthoryear{Fang, Noe, and Strack}{Fang
  et~al.}{2020}]{fang2020turning}
\textsc{Fang, D., T.~Noe, and P.~Strack} (2020): \enquote{Turning up the heat:
  The discouraging effect of competition in contests,} \emph{Journal of
  Political Economy}, 128, 1940--1975.

\bibitem[\protect\citeauthoryear{Fu and Wu}{Fu and Wu}{2019}]{fu2019contests}
\textsc{Fu, Q. and Z.~Wu} (2019): \enquote{Contests: Theory and topics,} in
  \emph{Oxford Research Encyclopedia of Economics and Finance}.

\bibitem[\protect\citeauthoryear{Glazer and Hassin}{Glazer and
  Hassin}{1988}]{glazer1988optimal}
\textsc{Glazer, A. and R.~Hassin} (1988): \enquote{Optimal contests,}
  \emph{Economic Inquiry}, 26, 133--143.

\bibitem[\protect\citeauthoryear{Immorlica, Stoddard, and Syrgkanis}{Immorlica
  et~al.}{2015}]{immorlica2015social}
\textsc{Immorlica, N., G.~Stoddard, and V.~Syrgkanis} (2015): \enquote{Social
  status and badge design,} in \emph{Proceedings of the 24th international
  conference on World Wide Web}, 473--483.

\bibitem[\protect\citeauthoryear{Konrad}{Konrad}{2009}]{konrad2009strategy}
\textsc{Konrad, K.~A.} (2009): \enquote{Strategy and dynamics in contests,} .

\bibitem[\protect\citeauthoryear{Krishna, Lychagin, Olszewski, Siegel, and
  Tergiman}{Krishna et~al.}{2024}]{krishna2024pareto}
\textsc{Krishna, K., S.~Lychagin, W.~Olszewski, R.~Siegel, and C.~Tergiman}
  (2024): \enquote{Pareto Improvements in the Contest for College Admissions,}
  .

\bibitem[\protect\citeauthoryear{Letina, Liu, and Netzer}{Letina
  et~al.}{2023}]{letina2023optimal}
\textsc{Letina, I., S.~Liu, and N.~Netzer} (2023): \enquote{Optimal contest
  design: Tuning the heat,} \emph{Journal of Economic Theory}, 105616.

\bibitem[\protect\citeauthoryear{Liu and Lu}{Liu and Lu}{2019}]{liu2019optimal}
\textsc{Liu, B. and J.~Lu} (2019): \enquote{The optimal allocation of prizes in
  contests with costly entry,} \emph{International Journal of Industrial
  Organization}, 66, 137--161.

\bibitem[\protect\citeauthoryear{Liu, Lu, Wang, and Zhang}{Liu
  et~al.}{2018}]{liu2018optimal}
\textsc{Liu, B., J.~Lu, R.~Wang, and J.~Zhang} (2018): \enquote{Optimal prize
  allocation in contests: The role of negative prizes,} \emph{Journal of
  Economic Theory}, 175, 291--317.

\bibitem[\protect\citeauthoryear{Liu and Lu}{Liu and Lu}{2017}]{liu2017optimal}
\textsc{Liu, X. and J.~Lu} (2017): \enquote{Optimal prize-rationing strategy in
  all-pay contests with incomplete information,} \emph{International Journal of
  Industrial Organization}, 50, 57--90.

\bibitem[\protect\citeauthoryear{Moldovanu and Sela}{Moldovanu and
  Sela}{2001}]{moldovanu2001optimal}
\textsc{Moldovanu, B. and A.~Sela} (2001): \enquote{The optimal allocation of
  prizes in contests,} \emph{American Economic Review}, 91, 542--558.

\bibitem[\protect\citeauthoryear{Moldovanu and Sela}{Moldovanu and
  Sela}{2006}]{moldovanu2006contest}
---\hspace{-.1pt}---\hspace{-.1pt}--- (2006): \enquote{Contest architecture,}
  \emph{Journal of Economic Theory}, 126, 70--96.

\bibitem[\protect\citeauthoryear{Moldovanu, Sela, and Shi}{Moldovanu
  et~al.}{2007}]{moldovanu2007contests}
\textsc{Moldovanu, B., A.~Sela, and X.~Shi} (2007): \enquote{Contests for
  status,} \emph{Journal of political Economy}, 115, 338--363.

\bibitem[\protect\citeauthoryear{Olszewski and Siegel}{Olszewski and
  Siegel}{2016}]{olszewski2016large}
\textsc{Olszewski, W. and R.~Siegel} (2016): \enquote{Large contests,}
  \emph{Econometrica}, 84, 835--854.

\bibitem[\protect\citeauthoryear{Olszewski and Siegel}{Olszewski and
  Siegel}{2020}]{olszewski2020performance}
---\hspace{-.1pt}---\hspace{-.1pt}--- (2020): \enquote{Performance-maximizing
  large contests,} \emph{Theoretical Economics}, 15, 57--88.

\bibitem[\protect\citeauthoryear{Onuchic and Ray}{Onuchic and
  Ray}{2023}]{onuchic2023conveying}
\textsc{Onuchic, P. and D.~Ray} (2023): \enquote{Conveying value via
  categories,} \emph{Theoretical Economics}, 18, 1407--1439.

\bibitem[\protect\citeauthoryear{Ostrovsky and Schwarz}{Ostrovsky and
  Schwarz}{2010}]{ostrovsky2010information}
\textsc{Ostrovsky, M. and M.~Schwarz} (2010): \enquote{Information disclosure
  and unraveling in matching markets,} \emph{American Economic Journal:
  Microeconomics}, 2, 34--63.

\bibitem[\protect\citeauthoryear{Popov and Bernhardt}{Popov and
  Bernhardt}{2013}]{popov2013university}
\textsc{Popov, S.~V. and D.~Bernhardt} (2013): \enquote{University competition,
  grading standards, and grade inflation,} \emph{Economic inquiry}, 51,
  1764--1778.

\bibitem[\protect\citeauthoryear{Rayo}{Rayo}{2013}]{rayo2013monopolistic}
\textsc{Rayo, L.} (2013): \enquote{Monopolistic signal provision,} \emph{The BE
  Journal of Theoretical Economics}, 13, 27--58.

\bibitem[\protect\citeauthoryear{Rodina and Farragut}{Rodina and
  Farragut}{2016}]{rodina2016inducing}
\textsc{Rodina, D. and J.~Farragut} (2016): \enquote{Inducing effort through
  grades,} Tech. rep., Tech. rep., Working paper.

\bibitem[\protect\citeauthoryear{Sisak}{Sisak}{2009}]{sisak2009multiple}
\textsc{Sisak, D.} (2009): \enquote{Multiple-prize contests--the optimal
  allocation of prizes,} \emph{Journal of Economic Surveys}, 23, 82--114.

\bibitem[\protect\citeauthoryear{Vojnovi{\'c}}{Vojnovi{\'c}}{2015}]{vojnovic2015contest}
\textsc{Vojnovi{\'c}, M.} (2015): \emph{Contest theory: Incentive mechanisms
  and ranking methods}, Cambridge University Press.

\bibitem[\protect\citeauthoryear{Zhang}{Zhang}{2024}]{zhang2024optimal}
\textsc{Zhang, M.} (2024): \enquote{Optimal contests with incomplete
  information and convex effort costs,} \emph{Theoretical Economics}, 19,
  95--129.

\bibitem[\protect\citeauthoryear{Zubrickas}{Zubrickas}{2015}]{zubrickas2015optimal}
\textsc{Zubrickas, R.} (2015): \enquote{Optimal grading,} \emph{International
  Economic Review}, 56, 751--776.

\end{thebibliography}

\end{document}